\documentclass[10pt,letterpaper]{article}
\usepackage[top=0.85in,left=1.50in,footskip=0.75in,marginparwidth=2in]{geometry}

\usepackage[utf8]{inputenc}

\usepackage{cite}

\usepackage{nameref,hyperref}

\usepackage[right]{lineno}
\usepackage{ltablex,booktabs}

\usepackage{microtype}
\DisableLigatures[f]{encoding = *, family = * }

\raggedright
\usepackage{changepage}

\usepackage[aboveskip=1pt,labelfont=bf,labelsep=period,singlelinecheck=off]{caption}

\makeatletter
\renewcommand{\@biblabel}[1]{\quad#1.}
\makeatother

\usepackage{lastpage,fancyhdr,graphicx}
\usepackage{epstopdf}
\fancyheadoffset[L]{2.25in}
\fancyfootoffset[L]{2.25in}

\usepackage{color}

\definecolor{Gray}{gray}{.25}

\usepackage{graphicx}

\usepackage{sidecap}

\usepackage{wrapfig}
\usepackage[pscoord]{eso-pic}
\usepackage[fulladjust]{marginnote}
\usepackage{longtable}

\reversemarginpar

\usepackage{enumitem}
\newlist{abbrv}{itemize}{1}
\setlist[abbrv,1]{label=,labelwidth=1in,align=parleft,itemsep=0.1\baselineskip,leftmargin=!}

\begin{document}
\nolinenumbers

\vspace*{0.35in}

% title goes here:
\begin{flushleft}
{\Large
\textbf\newline{PubTrend: General Overview of Artificial Intelligence for Colorectal cancer diagnosis from 2010–2022.}
}
\newline
% authors go here:
\\
Mary Adewunmi$^{1,3}$, Reem Abdel-Salam$^{2,3}$
\\
* mary.adewunmi@utas.edu.au, reem.abdelsalam13@gmail.com
\\
\bigskip
{1} University of Tasmania (UTAS), Hobart, Australia.
\\
{2} Cairo University, Egypt. 
\\
{3} CaresAI, Australia.
\bigskip
\\

\end{flushleft}

\section*{Abstract}
\justifying
Colorectal cancer (CRC) is among the most prevalent cancers in the world. Due to numerous scholarly papers and broad enquiries about specific use cases for artificial intelligence (AI) in colorectal cancer, researchers find it challenging to explore relevant papers on the current knowledge, comprehensive knowledge, and past methodologies in the literature review. This review extracts recent AI technology advances for diagnosing colorectal cancer from January 2010 to March 2022. PubTrends was used to identify and automate the intellectual structure and comparable papers on the use of AI in colorectal cancer diagnosis using the most cited papers, keywords, and similar papers.  Papers with quantitative results were represented with a tabular summary, and other paper contributions were in a sentence summary. Twenty-four (24) out of the forty-nine (49) top-cited papers were quantitative results, with one (1) outlier about lung cancer comprehensive screening. The most frequently used words were: "polyps,"  "detected", "image,"  and "colonoscopy." In addition, 83 per cent of the terms frequently used shortly before 2022 were image, polyps, detected, colonoscopy, and learning. In addition, 16 per cent are preparation, variant, classification, sample, and surgery.  The review showcases 49 of the 50 most cited papers, their notable contributions, objectives, specific AI methods, results, conclusions, and further recommendations. These papers highlight the limitations of colonoscopy for therapeutic use. The review concluded that despite the enormous benefits of using artificial intelligence, from improving diagnosis, the medical AI programmer still needs to be actively involved in the diagnosis team for effective results in CRC diagnosis.

\subsection*{Keywords}\mbox{PubTrend, AI, CRC, colonoscopy, diagnosis}

% now start line numbers
\nolinenumbers

% the * after section prevents numbering
\section*{Introduction}

Colorectal cancer (CRC) is a significant human cancer with a high mortality rate, accounting for approximately 1.5 million newly diagnosed cases and 500,000 deaths in 2022, according to the cancer statistics field \cite{siegel2019cancer}. Artificial intelligence with a colorectal cancer diagnosis was explored using the Pubtrend library and further analysed using a study field's intellectual structure and comparable papers. https://pubtrends.net uses PubMed biomedical papers. The library uses bibliometrics approaches for citation information analysis and natural language processing algorithms to compute paper similarity, then topic extraction and clustering. integrated citation graph and article similarity network viewer with powerful visualisation and filtering options simplify research area exploration. 
Finally, deep learning generates interactive literature evaluations. This review summarises the recent advances in colorectal cancer diagnosis using artificial intelligence.

\section*{Methods}

PubTrends is a real-time PubMed scraper for extracting biomedical papers, https://pubtrends.net \cite{shpynov2021pubtrends}. It analyses a study field's intellectual structure and comparable papers using the PubMed bibliographic database \cite{canese2013pubmed} for citation information analysis. Natural language processing algorithms compute paper similarity, followed by topic extraction and clustering. Additionally, it displays the most cited papers, keywords, authors, and journals in search results. The 50 most cited papers were utilised to structure the review into tabular and sentence summaries. This includes papers with quantitative analysis results, sentence summaries, reviews, most frequent words, similar reports, and paper insights.

\section*{Results}
The results were structured into tabular summaries (see Appendix A), which include the papers with quantitative analysis results, sentence summaries, papers with reviews, most frequent words in the study, most similar reports, and paper insights. 

\subsection*{Sentence-based Summary}
While researching colorectal cancer drugs in 2010, Martinez looked at the latest ontology and complex network studies. He also looked at the Gene Ontology's drug metabolic process sub-ontology topology \cite{martinez2010artificial}. Optical coherence tomography, confocal endomicroscopy, narrow-band imaging, autofluorescence imaging, virtual chromoendoscopy, and volumetric laser endomicroscopy are some of the new imaging techniques that scientists are using to look inside the digestive tract and figure out what is wrong. Computer-aided diagnosis (CAD) systems may make it easier for multiple observers to find and evaluate mucosal lesions. This could lead to a 20 per cent miss rate for colon polyps and a few colorectal cancers found after a colonoscopy \cite{de2018methodology}. Additionally, Ahmad et al. (2019) stressed the importance of CAD for reducing the number of missed polyps and interval colorectal cancers after colonoscopy, which was 22 per cent.  Artificial intelligence in gastroenterology can be achieved by focusing on endoscopic-based autonomous diagnosis \cite{min2019overview}. A 3 per cent decrease in interval CRC incidence has been linked to a 1 per cent rise in the adenoma detection rate \cite{azer2019challenges }.  Convolutional neural networks (CNNs) in deep learning attempts are used to detect colonic polyps and malignant lesions, considering these limitations. CNN has improved polyps identification, segmentation and categorisation with improved and accurate diagnosis \cite{azer2019challenges}. The most recent and the state of the art paper on uses computer aided diagnostics to help diagnose Barrett's oesophagus and early oesophageal squamous cell carcinoma, as well as predicting the depth of invasion in oesophageal tumours using both standard and cutting-edge endoscopic techniques\cite{lazuar2020impact}. The limitations of AI-assisted endoscopy for effective application in clinical usage are mostly common with new technologies \cite{abadir2020artificial}. (Acs et.al (2020) showed recent advances in digital image analysis and diagnostic pathology \cite{acs2020artificial}. The progress made in science and the real-world use of AI in cancer research, early diagnosis, treatment, and prognosis showed a theoretical framework for how AI could be used as a diagnostic and treatment tool for cancer \cite{wang2022time}. Computer-assisted optical biopsies can find low-risk polyps that can be removed and thrown away or diagnosed and left alone, which saves money that would otherwise be spent on unnecessary procedures \cite{shung2020artificial, sinagra2020use}. Namikawa et al. (2020) suggested the expert opinion of endoscopists on the application of AI for carrying out endoscopy operations \cite{namikawa2020utilizing, wang2022cost}. AI had a higher average area under the receiver operating characteristics curve (AUC) than pathologists (0.988 vs 0.970) and performed best among AI approaches for CRC diagnosis. Machine learning models could revolutionise medicine and individual cancer therapy. Integrating AI in colonoscopy and modern endoscopic modalities such as magnifying narrow-band imaging, endocytoscopy, confocal endomicroscopy, laser-induced fluorescence spectroscopy, and magnifying chromoendoscopy can improve polyp identification and characterisation \cite{goyal2020scope}. The ongoing progress of AI for colon capsule endoscopy for colorectal cancer screening and its potential use in inflammatory bowel disease \cite{hosoe2018establishment}. Artificial intelligence improves quality and reduces costs in colonoscopy-based colorectal screening and monitoring. Real-time computer-assisted polyp detection improves adenoma detection rates.  AI was used to study the upper, middle, and lower gastrointestinal tracts, inflammatory bowel disease, the hepatobiliary system, and the pancreas \cite{kroner2021artificial}. The results showed the clinical applications of AI, its limitations, and future directions in this field. Computer-aided detection and characterisation technologies enhance adenoma detection. Another study proposed robotic surgery and computer-assisted drug delivery to advance CRC treatment with precision or individualised medicine \cite{mitsala2021artificial}. AI proposes a self-attention-based YOLOv5 model for polyp target identification. Regression is used to feed the network the whole picture and return the target frame of this point in multiple image positions. An attention method is introduced to feature extraction to boost information-rich feature channels and reduce unnecessary channel interference.  The newest research and methods to find problems in the gut using machine learning are based on DNA and microbiome data that is not attached to cells and is generated by amplicon or whole-genome sequencing \cite{zhang2021noninvasive}. A look at the first 500 videos that used the current state of CADe research, the problems with these kinds of systems, and legal issues related to AI performance for preclinical testing \cite{fitting2022video}. Discovery of how AI will affect colorectal cancer epidemiology and new mass information-collecting approaches like GeoAI, digital epidemiology, and real-time information collection \cite{yu2022role}. Deep learning also improved CT/MRI, endoscopes, genomics, and pathology evaluations. Finally, AI may improve CRC therapy. AI's therapeutic prescription for colorectal cancer shows potential in clinical and translational oncology, which promises better, more customised therapies for patients. Significant studies highlight the limitations of CADe and its clinical acceptance for colonoscopy \cite{oliveira2021cad}. In the end, AI can diagnose and explain for colorectal cancer in oncologic imaging \cite{vicini2022narrative}.

\begin{figure}[] %s state preferences regarding figure placement here

% use to correct figure counter if necessary
%\renewcommand{\thefigure}{2}

\includegraphics[width=\textwidth]{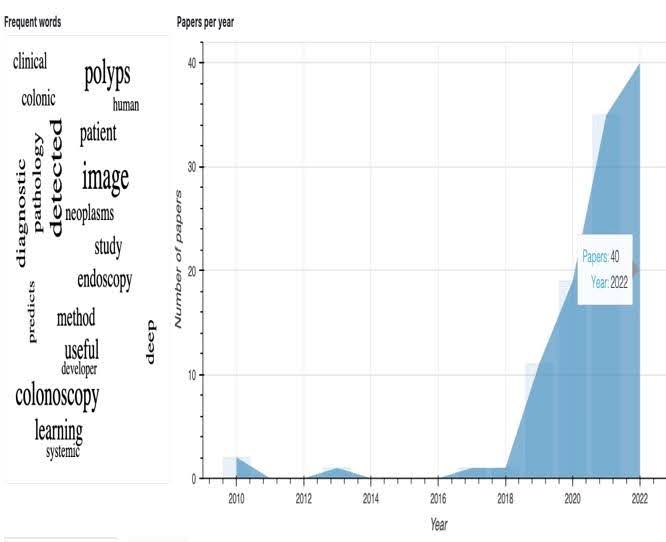}

\caption{\color{Gray} \textbf{Right: Most Frequent words from the 50 most cited papers. Left: Figure 1b. Papers with similar words, references, and co-citations.}}

\label{fig1} % \label works only AFTER \caption within figure environment

\end{figure}

\subsection*{Most Frequent Words}
As shown in Figure 1a, out of the 50 topmost cited papers from 2010 to 2022, the most frequent words are polyps, detected, image and colonoscopy. Figure 1b shows forty (40) papers in 2022 with topics with similar ideas, descriptions, etc. Topics with similar references, co-citations, direct citations, and text were extracted. Image, polyps, detected, colonoscopy, and learning made up 83 per cent of the words used. Preparation, variant, classify, sample, and surgery made up 16 per cent. These words were used more often before 2022 compared to learning, neoplasms, and the decline in useful, human, learning, method, predicts, patient, study, pathology, amongst and clinical, image, colonoscopy, systemic, polyps, and endoscopy.

\begin{figure}[] %s state preferences regarding figure placement here

% use to correct figure counter if necessary
%\renewcommand{\thefigure}{2}

\includegraphics[width=\textwidth]{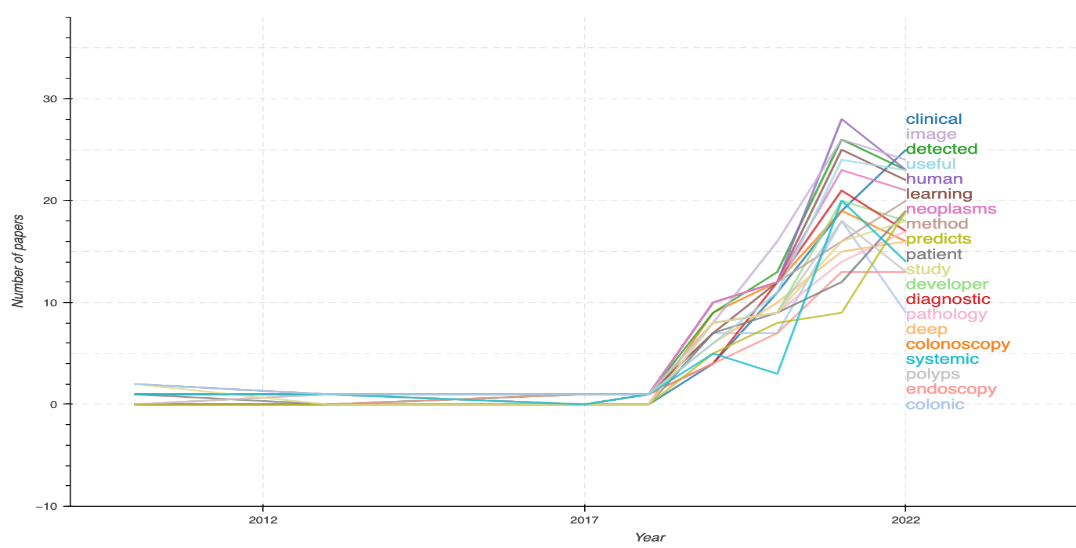}

\caption{\color{Gray} \textbf{Keyword frequencies of top-cited papers.}}

\label{fig2} % \label works only AFTER \caption within figure environment

\end{figure}

\subsection*{Most Similar Paper}

\begin{figure}[] %s state preferences regarding figure placement here

% use to correct figure counter if necessary
%\renewcommand{\thefigure}{2}

\includegraphics[width=\textwidth]{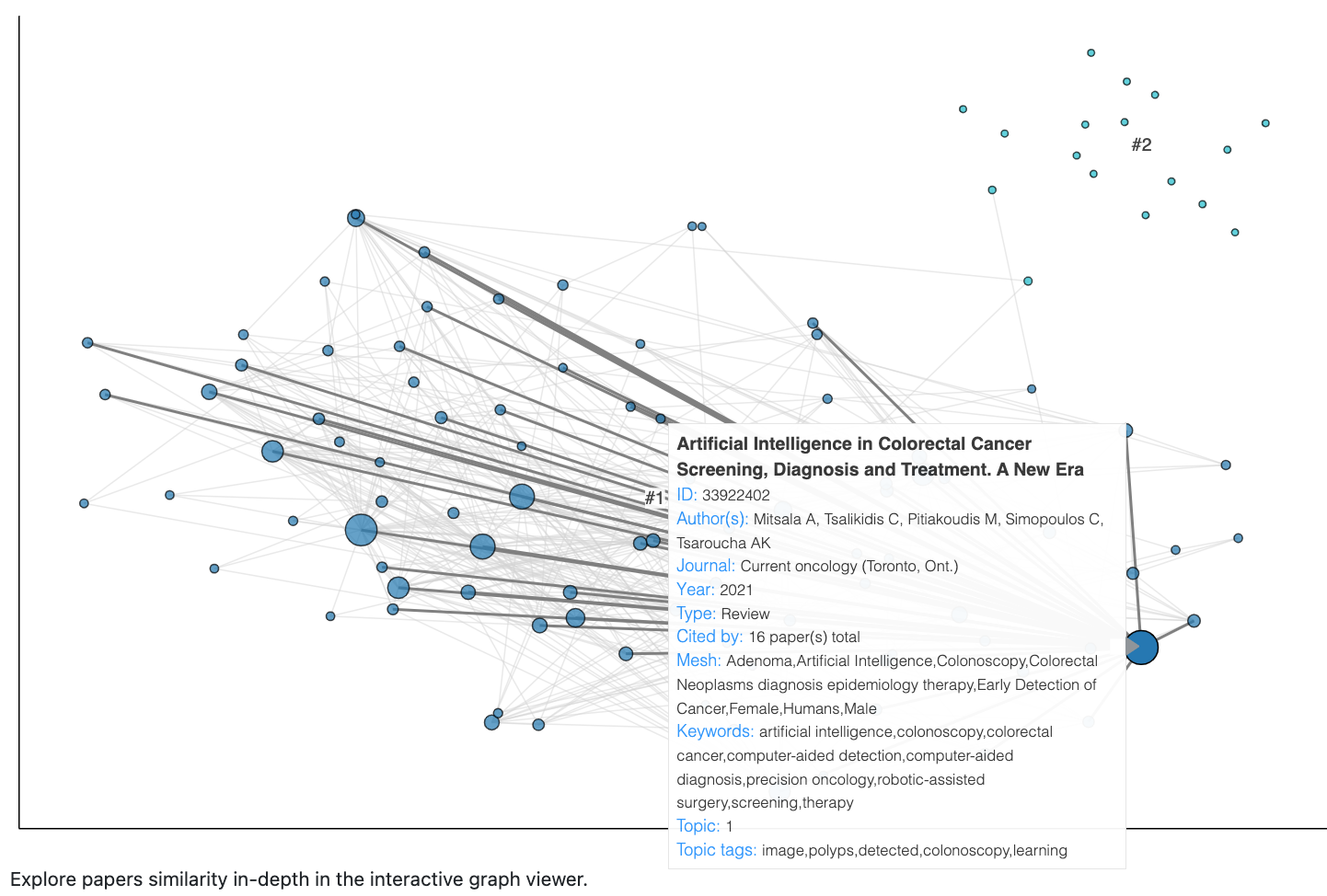}

\caption{\color{Gray} \textbf{Graphical Network analysis showing the similarity of papers with the theme (AI in colorectal cancer diagnosis).}}

\label{fig3} % \label works only AFTER \caption within figure environment

\end{figure}

The highest correlated paper with the theme was Artificial Intelligence in Colorectal Cancer by Mitsala et al.(2021) as shown in Figure \ref{fig3}. The paper examines AI applications in CRC screening, diagnosis, and therapy and their promising outcomes in improving adenoma detection, such as AI-assisted screening, computer-aided detection, and characterisation techniques. The article further noted that robotic surgery and computer-assisted medication delivery could advance CRC therapy. 

\section*{Conclusion}
In this review, the most cited top fifty (50) were analysed with twenty-four (24) quantitative research works, summarised in tabular form and twenty-five (25) qualitative reviews in sentence-base form.  One (1) outlier in lung cancer screening was ignored due to its irrelevancy to the research theme. The most frequent words from 2010 to 2022 were polyps, detected, colonoscopy, and learning, and 16 per cent were preparation, variant, classify sample and surgery. This insight explains why colonoscopy is still the primary standard for CRC polyps screening, but could be assisted with model learning and classification. This paper also emphasised ways that AI and machine or deep learning can improve CT/MRI, endoscopes, genomics, pathology evaluations and CRC therapy. AI's therapeutic prescription for colorectal cancer shows potential in clinical and translational oncology, which promises better, more customised therapies for patients. Significant studies highlight the limitations of colonoscopy for therapeutic use. Notwithstanding the many benefits of using artificial intelligence, from improving diagnosis to reducing diagnosis costs and assisting in precision medicine, the medical AI programmer still needs to be actively involved in the diagnosis for effective and successful results.

\nolinenumbers

%This is where your bibliography is generated. Make sure that your .bib file is actually called library.bib
\bibliography{library}

%This defines the bibliographies style. Search online for a list of available styles.
\bibliographystyle{ieeetr}

\section{Appendices}
\subsection{Appendix A}
\setlength\LTleft{-1.5in}
\setlength\LTright{0pt}
\captionof{table}{Tabular summary review}
\begin{longtable}{p{1.85cm}p{2.5cm}p{2cm}p{3cm}p{4cm}p{2cm}}
 
         Author /Year&  Topic&  Objectives&  Methods&  Results& Conclusion/Future Recocmmendation\\
         (Glissen Brown et al., 2022)

\cite{wang2019real}
         &

         Real-time automatic detection system increases colonoscopic polyp and adenoma detection rates: a prospective randomised controlled study & evaluation of a deep-learning-based CADe system in a prospective, multicentre, single-blind, randomised tandem colonoscopy investigation (EndoScreener, Shanghai Vision AI, China).
&  232 participants were randomised to have CADe or HDWL colonoscopy first. 223 patients after 9 were excluded.
&  CADe-first patients had a lower AMR (20.12 per cent [34/169] vs 31.25 per cent [45/144]; OR, 1.8048; 95 per cent CI, 1.0780–3.0217; P =.0247) than HDWL-first patients. CADe-first had a lower SSL miss rate (7.14 per cent [1/14]) than HDWL-first (42.11 per cent [8/19]; P=.0482). CADe-first group first-pass APC was more significant (1.19 [SD, 2.03] vs 0.90 [SD, 1.55]; P=.0323). First-pass ADR was 50.44 per cent in the CADe-first group and 43.64 per cent in HDWL-first (P=.3091).
& CADe-systems reduce AMR and SSL miss rates and boost first-pass APC compared to HDWL colonoscopy.
\\
      (Yamada et al., 2019)

   \cite{yamada2019development}   & Development of a real-time endoscopic image diagnosis support system using deep learning technology in colonoscopy  &  Endoscopists' colonoscopy skills are lacking owing to experience, and remedies are required. Thus, a real-time, robust colorectal neoplasm detection system reduces the likelihood of colonoscopy-missed lesions. Developed an artificial intelligence (AI) system that automatically detects early signs of colorectal cancer during colonoscopy.
&  Colonoscopy
&  The AI system's sensitivity and specificity are 97.3\% (95\% CI = 95.9\%–98.4\%) and 99.0\% (95\% CI = 98.6\%–99.2\%), respectively, and the validation set's area under the curve is 0.975 (95\% CI = 0.964–0.986). The sensitivities are 98.0\% (95\% CI = 96.6\%–98.8\%) in the polypoid subgroup and 93.7\% (95\% CI = 87.6\%–96.9\%) in the non-polypoid. 
& Tensor metrics in the trained model were deconstructed to speed detection, and the system can identify malignant areas 21.9 ms/image on average
\\
      (Thakur, Yoon, & Chong, 2020) \cite{thakur2020current}  &  Current trends of Artificial Intelligence for Colorectal Cancer Pathology Image Analysis: A Systematic Review \&  To conduct a comprehensive assessment of AI applications in CRC image pathology analysis
&  Searched online databases, including MEDLINE, for studies published between January 2000 and January 2020. (PubMed, Cochrane Library, and EMBASE). "Colorectal neoplasm," "histology," and "artificial intelligence" were among the search terms. From the 9000 recognised studies, 30 research papers with 40 models were chosen for review
&  The models' algorithm characteristics included gland segmentation (n = 25, 62\%), tumour classification (n = 8, 20\%), tumour microenvironment characterisation (n = 4, 10\%), and prognosis prediction (n = 3, 8\%). Only 20 gland segmentation models passed the quantitative analysis criteria, and Ding et al. (2019) model performed the best.
& Validation at the level of routine practice requires future investigations with larger datasets and high-quality annotations with a promising CRC pathological analysis.
\\
     (K.-S. Wang et al., 2021)\cite{wang2021accurate}   & Accurate diagnosis of colorectal cancer based on histopathology images using artificial intelligence
 &  To reduce clinical pathologist’s bias of tiredness thereby speeding up CRC diagnosis
&  A novel patch aggregation technique for clinic CRC diagnosis employs weakly labelled pathological whole-slide image (WSI) patches and a state-of-the-art transfer-learned deep convolutional neural network in AI. 170,099 patches, 14,680 WSIs, and 9631 subjects from China, the US, and Germany were used to train and test this technique
&  The model accurately diagnosed CRC WSIs from multicenters (average Kappa statistic 0.896) and often outperformed most expert pathologists. AI had a higher average area under the receiver operating characteristics curve (AUC) than pathologists (0.988 vs 0.970)
& This first generalisable AI system can reliably process vast numbers of WSIs without clinical pathologists' tiredness bias. It will greatly reduce the burden of everyday pathology diagnosis and improve CRC treatment.
\\
   (Kel et al., 2019)  \cite{kel2019walking}    & Walking pathways with positive feedback loops reveal DNA methylation biomarkers of colorectal cancer
  &  To find out novel bioinformatic approaches required for multi-omics data analysis and to identify causal biomarkers that may drive early cancer.
  
& Devised a technique to discover probable causal links between epigenetic modifications (DNA methylations) in gene regulatory areas that impact transcription factor binding sites (TFBS) and gene expression changes. This approach also examines signal transduction pathway structure and looks for positive feedback loops that may generate cancerous gene expression abnormalities

 &  Analysed an extensive collection of full genome gene-expression data (RNA-seq) and DNA methylation data of genomic CpG islands (using Illumina methylation arrays) from a sample of tumour and normal gut epithelial tissues from 300 colorectal cancer patients at different stages of the disease (data generated in the EU-supported SysCol project). MGE's automated multi-omics analysis online service identified possible epigenetic biomarkers of DNA methylation (my-genome-enhancer.com). MGE analyses cancer-specific enhancers using TRANSFAC®, TRANSPATH®, and AI-based technologies.

& Tested biomarkers on an independent collection of colorectal cancer patient blood samples. Thus, utilising advanced statistics and machine learning, a minimal set of 6 biomarkers was chosen to maximise cancer diagnosis. CALCA, ENO1, MYC, PDX1, TCF7, and ZNF43 have hypermethylated regulatory areas
\\

        (Q. Wang et al., 2019) \cite{wang2019establishment} & Establishment of multiple diagnosis models for colorectal cancer with artificial neural networks

 & To develop various colorectal cancer (CRC) detection models using data from The Cancer Genome Atlas (TCGA) database and artificial neural networks for improving CRC diagnosis.

 & A genetic algorithm and mean effect value select genes to encode numerical parameters for cancer metastasis or hostility. Cancer/Normal, M0/M1, carcinoembryonic antigen (CEA) <5/5, and Clinical stage I/II/III/IV were diagnosed using backpropagation and learning vector quantization neural networks. AUC, and a 10-fold cross-validation test. The 

 &  Each model was assessed using prediction accuracy (ACC). Cancer/Normal, M0/M1, CEA, and Clinical stage models have 100\% ACC and AUC. 93.75–99.39\%, 1.0000; 80.58–88.24\%, 0.9286–1.0000; 67.21–92.31\%, 0.7091–1.0000; and 59.13–68.85\%, 0.6017–0.6585. 
& This work-built CRC diagnostic models uses gene expression profiling data and artificial intelligence algorithms.
\\
  (Dong et al., 2020) \cite{dong2020clinical}      & Clinical Trials for Artificial Intelligence in Cancer Diagnosis: A Cross-Sectional Study of Registered Trials in ClinicalTrials.gov

 & evaluate AI cancer diagnostic experiments.

 & ClinicalTrials.gov was searched and downloaded for AI cancer diagnostic studies. SPSS 20.0 analysed data.

 &97 trials were recorded. 27 were interventional and 70 were observational. 15 (15.4\%) experiments were completed. 18 unrecruited trials and 50 in recruitment. 31 (32.0\%) trials contained 100–499 cases, while 17 (17.5\%) covered 500–999 instances. Only two of 27 interventional studies reported phase. Interventional trials were mainly used for cancer diagnosis (85.2\%) and treatment (3.7\%). 46 (65.7\%) observational clinical trials were cohort studies, and 11 (15.7\%) were case-only studies. 46 (65.7\%) observational trials were prospective, and 13 (18.6\%) retrospectives. 37 (38.1\%) of 97 studies covered colorectal cancer, 11 (11.3\%) breast cancer, 43 (44.3\%) imaging diagnosis, 33 (34.0\%) endoscopic diagnosis, and 11 (11.3\%) pathological diagnosis. 11 interventional studies were parallel (40.7\%) and 14 were single group (51.9\%). 18 (66.7\%) of the 27 interventional studies had no masking, 6 (22.2\%) had single masking, 1 (3.7\%) had double masking, and 2 (7.4\%) had triple masking.

  &Most AI cancer detection experiments are observational, and more are required

 \\
    (Nazarian, Glover, Ashrafian, Darzi, \& Teare, 2021)\cite{nazarian2021diagnostic}  & Diagnostic Accuracy of Artificial Intelligence and Computer-Aided Diagnosis for the Detection and Characterization of Colorectal Polyps: Systematic Review and Meta-analysis

 &This review assessed AI-based colorectal polyp diagnostics’ accuracy.

  & Embase, MEDLINE, and the Cochrane Library were used to search the literature. PRISMA standards.

 & 48 studies were included. The meta-analysis indicated that AI-detected polyps had a significantly higher pooled polyp detection rate than conventional colonoscopy (OR 1.75, 95\% CI 1.56-1.96; P<.001). ADR was also higher in AI-treated colonoscopy patients (OR 1.53, 95\% CI 1.32-1.77; P<.001)

 & Machine learning can enhance ADR and lower CRC. AI-based colorectal polyp identification and characterisation methods are very accurate

\\
      (Liew, Tang, Lin, \& Lu, 2021)  \cite{liew2021automatic} & Automatic colonic polyp detection using the integration of modified deep residual convolutional neural network and ensemble learning approaches

  &This work develops a new CAD tool to identify colonic polyps reliably.

  & Modified deep residual network to discriminate colonic polyps, principal component analysis, and AdaBoost ensemble learning. Altering ResNet-50, a strong deep residual network design, reduced computing time. The classification model was trained on endoscopic images using a median filter, picture thresholding, contrast enhancement, and normalisation to minimise interference. The model was trained using images with and without polyps from Kvasir, ETIS-LaribPolypDB, and CVC-ClinicDB, three publically accessible datasets.

 & The suggested technique trained with three datasets has an MCC of 0.9819, accuracy, sensitivity, precision, and specificity of 99.10\%, 98.82\%, 99.37\%, and 99.38\%, respectively.
 & Early identification of endoscopic pictures using computer-aided diagnostics tools

 \\
     (Xu et al., 2021)(Wallace et al., 2022) \cite{xu2021comparison}    & Comparison of diagnostic performance between convolutional neural networks and human endoscopists for diagnosis of colorectal polyp: A systematic review and meta-analysis

 & To check whether the CNN system has significant drawbacks and to confirm whether it outperforms human endoscopists

 & Studies from April 2020 were searched in PubMed, Web of Science, Cochrane Library, and EMBASE. The enrolled studies were also assessed using QADAS-2. Deeks' funnel plot determined publication bias. The meta-analysis included 13 studies (ranging between 2016 and 2020). 

 & CNN system performed well in CP detection (sensitivity: 0.848 [95\% CI: 0.692–0.932]; specificity: 0.965 [95\% CI: 0.946–0.977]; and AUC: 0.98 [95\% CI: 0.96–0.99]) and classification (sensitivity: 0.943 [95\%: 0.927–0.955]; specificity: 0.894 [95\%: 0.631–0.977]; and AUC: 0.95 [95\%: 0.93–0.97]).

 &CNN system performed similarly to the expert but much better than the non-expert in CP classification (CNN vs expert: RDOR: 1.03, P = 0.9654; non-expert vs expert: 0.29, P = 0.0559; non-expert vs CNN: 0.18, P = 0.0342). Thus, the CNN method performed well for CP diagnosis and could be employed during colonoscopy.

 \\
   (Wallace et al., 2022)  \cite{wallace2022impact}     & Impact of Artificial Intelligence on Miss Rate of Colorectal Neoplasia

 & To improve colorectal neoplasia identification and CRC prevention.

 & In 8 sites (Italy, UK, US), CRC screening or surveillance patients were randomised (1:1) to receive two same-day, back-to-back colonoscopies with or without AI (deep learning computer-aided diagnostic endoscopy) in two arms: AI followed by colonoscopy without AI or vice-versa. The adenoma miss rate (AMR) was estimated by dividing the histologically validated lesions found during second colonoscopy by the total number seen at both colonoscopies. The mean number of lesions found in the second colonoscopy and the fraction of false negatives (no lesion at first and at least 1 at second) were determined. Endoscopist, age, sex, and colonoscopy indication adjusted ORs and 95\% CIs. Measured adverse incidents.

 & 230 individuals (116 AI first, 114 conventional colonoscopies first) were analysed. AMR was 15.5\% (38 of 246) and 32.4\% (80 of 247) in the AI and non-AI colonoscopy first arms, respectively (adjusted OR, 0.38; 95\% CI, 0.23–0.62). AMR was lower for AI in the proximal (18.3\% vs 32.5\%; OR, 0.46; 95\% CI, 0.26–0.78) and distal colon (10.8\% vs 32.1\%; OR, 0.25; 95\%, 0.11–0.57) colons. AI-first colonoscopy had fewer adenomas at second colonoscopy (0.33 ± 0.63 vs 0.70 ± 0.97, P <.001). The AI and non-AI first arms had 6.8\% (3 of 44) and 29.6\% (13 of 44) false negative rates, respectively (OR, 0.17; 95\% CI, 0.05–0.67). Adverse occurrences were similar across groups.

 & AI reduced colorectal neoplasia miss rate by around 2-fold, demonstrating its use in minimising perceptual mistakes for tiny and subtle lesions during a routine colonoscopy.

\\

   (Shi, Su, Zhang, Huang, \& Zhu, 2010) \cite{shi2010intelligent}     &An intelligent decision support algorithm for diagnosis of colorectal cancer through serum tumor markers

  & Optimizing check combinations and maximising check values improves diagnostic accuracy (DA) and lowers cost

 & This research develops DS-STM, an artificial intelligence system for colorectal cancer (CRC). DS-STM helps doctors choose tumour markers and diagnose colorectal cancer (CRC). Most CRC patients only need two tumour markers, according to the research

 & DS-STM improved DA from 67.53\% to 73.87\% for the same validation dataset compared to the serial test. 

 &The new diagnosing technique also cuts costs.

 \\
   (Gupta et al., 2021) \cite{gupta2021colon}     &Colon Tissues Classification and Localization in Whole Slide Images Using Deep Learning

  &Find the region of the colorectal tissues that is normal and pathological in a quicker, more precise, and more consistent manner.

  & Pretrained Inception-v3 model

 & With the pre-trained Inception-v3 model, the proposed models had an F-score of 0.97 and an area under the curve (AUC) of 0.97. With the customised Inception-ResNet-v2 Type 5 (IR-v2 Type 5) model, the F-score and AUC were both 0.99

 & \\
     (Pellegrino et al., 2021) \cite{pellegrino2021machine}    &  Machine learning random forest for predicting oncosomatic variant NGS analysis

 & Presented a machine learning approach for categorising pathogenic single nucleotide variants (SNVs), single nucleotide polymorphisms (SNPs), multiple nucleotide variants (MNVs), insertions, and deletions discovered by NGS from various tumour specimen types, including colorectal, melanoma, lung, and glioma cancer

  & Evaluated the performance of the various ML algorithms and determined whether one is a suitable model for verifying NGS variant calls in cancer diagnosis by comparing NGS data to several ML algorithms using the k-fold cross-validation method and to neural networks (deep learning)

 & Machine learning with 70\% of data samples, retrieved from local database and validated it with the remaining 30\% of data. The NGS analysis method was constructed using the model with the highest accuracy. Version 3.6.0 of the R scripting language was used to create artificial intelligence. 102,011 variants, or 70\%, to train model. With random forest machine learning (ntree = 500 and mtry = 4), the best error rate (0.22\%), and the AUC was 0.99. Positive results were achieved by neural networks. With validation data, the final trained neural network model has an accuracy of 98\% and a ROC-AUC of 0.99. More than 2000 variants from NGS database were used to test RF model's interpretation, and 20 variants were incorrectly categorised (error rate, 1\%). Error rate was consistently less than 0.5\% after regularly executing RF model and adding false positives to the training database

 & The RF model is simple to use in other molecular biology labs and provides great results for oncosomatic NGS interpretation and neural networks may be helpful in anticipating more complicated variants because they now have strong capability in variant categorisation

\\
   (Deding et al., 2020) \cite{deding2020colon}      & Colon capsule endoscopy versus CT colonography after incomplete colonoscopy. Application of artificial intelligence algorithms to identify complete colonic investigations.

 & To build a forward-tracking algorithm for completion of combined investigations, investigate the relative sensitivity of CCE compared with CTC following incomplete OC, and investigate the completion rate when combining results from the incomplete OC and CCE.

 & Patients having criteria for CTC after incomplete OC were included for CCE and CTC in this prospective paired trial. To identify comprehensive integrated investigations, the locations of CCE and OC abortions were registered. An AI-based system for capsule localisation was created by recreating the colon's path

 & A total of 105 people—97 of whom underwent both a CCE and a CTC—were included in the 237 people with CTC indication. The CCE was finished in 66 (68\%). Including CCEs that had progressed to the most advanced stage of incomplete OC, 73 (75\%) had finished their colonic investigations, and 78 (80\%) had concluded. For polyps larger than 5 mm and larger than 9 mm, the relative sensitivity of CCE compared to CTC was 2.67 (95\% confidence interval (CI) 1.76;4.04) and 1.91 (95\% CI 1.18;3.09), respectively

  & Introducing and upgrading algorithm-based localisation of capsule abortion may boost the detection of overall complete investigation rates following incomplete OC. The sensitivity of CCE following incomplete OC was superior to CTC.

\\
    (Bedrikovetski et al., 2021) \cite{bedrikovetski2021artificial}     & Artificial intelligence for pre-operative lymph node staging in colorectal cancer: a systematic review and meta-analysis

 & To assess the diagnostic reliability of AI models utilised for the early detection of lymph node metastases in colorectal cancer staging images

 &A literature search of research published from January 2010 to October 2020 using PubMed (MEDLINE), EMBASE, IEEE Xplore, and the Cochrane Library was done by PRISMA standards. Included were studies that examined the precision of radiomics models and/or deep learning for the CT/MRI detection of lymph node metastases in colorectal cancer. Abstracts from conferences and studies that focused on image segmentation accuracy rather than nodal classification were removed. The studies' quality was evaluated with the help of a customised questionnaire based on the QUADAS-2 criteria, using the study's characteristics and diagnostic metrics taken. The area under the receiver operating characteristic curve (AUROC) was pooled in a meta-analysis.

  & For the systematic review, 17 studies that met the criteria were found, of which 12 used radiomics models and five deep learning models. Two studies had a high risk of bias, and radiomics publications had significant heterogeneity (73.0\%). For radiomics and deep learning models, the per-patient AUROC for rectal cancer was 0.808 (0.739-0.876) and 0.917 (0.882-0.952), respectively. With an AUROC of 0.688, the radiologists underperformed both models (0.603 to 0.772). In the same manner, radiomics models with a per-patient AUROC of 0.727 (0.633-0.821) outperformed radiologist models with an AUROC of 0.676 in colorectal cancer (0.627–0.725)

 & Although studies on radiomics and deep learning are sparse, AI models can predict lymph node metastases in rectal and colorectal cancer more effectively

\\
     (Brown et al., 2022) \cite{brown2022deep}   & Deep Learning Computer-aided Polyp Detection Reduces Adenoma Miss Rate: A United States Multi-Center Randomized Tandem Colonoscopy Study (CADeT-CS Trial)

 & To check the impact of CADe, the problem of missing polyps during colonoscopy in a population in the United States (U.S.) during screening and surveillance colonoscopies.

 &Multicenter, single-blind, randomised tandem colonoscopy investigation (EndoScreener, Shanghai Wision AI, China) was adopted. From 2019 to 2020, patients were enrolled at 4 academic medical centres in the United States. The same endoscopist randomly assigned patients presenting for colorectal cancer screening or surveillance to either a CADe colonoscopy or a high-definition white light (HDWL) colonoscopy first, then the other operation in tandem right after. Sessile serrated lesion (SSL) miss rate, adenomas per colonoscopy, and adenoma miss rate (AMR) were the secondary outcomes (APC).

  & A total of 232 participants were enrolled in the trial, and 116 of them were randomly assigned to either a CADe colonoscopy or an HDWL colonoscopy initially. Then, 9 patients were removed from the study cohort, leaving 223 patients remaining. Compared to the HDWL-first group, the CADe-first group had a lower AMR (20.12\% [34/169] vs. 31.25\% [45/144]; odds ratio [OR], 1.8048; 95\% confidence interval [CI], 1.0780-3.0217; P =.0247). SSL miss rate in the CADe-first group was lower (7.14\% [1/14] vs. 42.11\% [8/19]; P =.0482) than in the HDWL-first group. In the CADe-first group, first-pass APC was greater (1.19 [standard deviation (SD), 2.03] vs. 0.90 [SD, 1.55]; P =.0323). In the CADe-first group, first-pass ADR was 50.44\%, whereas in the HDWL-first group, it was 43.64\% (P =.3091).

 &In tandem colonoscopy randomised controlled experiment, CADe-system in comparison to HDWL colonoscopy alone reduces AMR and SSL miss rate and increases first-pass APC

 \\
   (Gao, Guo, Sun, \& Qu, 2020)  \cite{gao2020application}    & Application of Deep Learning for Early Screening of Colorectal Precancerous Lesions under White Light Endoscopy

 & To develop colorectal lesion detection, positioning, and classification models based on white light endoscopic images using deep learning techniques, as well as to design a computer-aided diagnosis (CAD) system to assist physicians in lowering the rate of missed diagnoses and raising the detection rate's accuracy

 & white light endoscopic pictures taken while certain patients were having colonoscopies. The convolutional neural network model determines the presence of CRC, colorectal adenoma (CRA), and colorectal polyps. The model was assessed using the rates of sensitivity, specificity, and accuracy. The lesions on the images with lesions are then located and classified using the instance segmentation model, and the performance of an instance segmentation model is assessed using the metrics mAP (mean average precision), AP50, and AP75

 & Use ResNet50 to the other four models—AlexNet, VGG19, ResNet18, and GoogLeNet—to determine whether the image has lesions. As a result, ResNet50 outperforms several other models. It received a 93.0\% accuracy rating, a 94.3\% sensitivity rating, and a 90.6\% specificity rating. The mAP, AP50, and AP75 of the lesion used in the localisation and classification of the lesion by Mask R-CNN were 0.676, 0.903, and 0.833, respectively.

 &Mask R-CNN model may be used to find and categorise lesions in images with lesions, and ResNet50 demonstrated the best performance.

 \\
   (Horiuchi et al., 2019) \cite{horiuchi2019real}     & Real-time computer-aided diagnosis of diminutive rectosigmoid polyps using an auto-fluorescence imaging system and novel color intensity analysis software

 & To create software that can achieve $\geq$90\% NPV, distinguish between rectosigmoid tiny polyps by using the green-to-red (G/R) ratio,

 & Patients who were scheduled for endoscopic treatment at facility from December 2017 to May 2018 and had known polyps were prospectively enrolled. First, computer-aided diagnosis with autofluorescence imaging (CAD-AFI) was used to distinguish between all colorectal diminutive polyps using a novel software-based automatic colour intensity analysis. Endoscopists then made a diagnosis based on the results of trimodal imaging endoscopy (TME), which combines AFI, white-light imaging (WLI), and magnifying narrow-band imaging (M-NBI). Following the endoscopic removal of all polyps, the histological diagnosis was assessed.

 &95 patients with 258 small rectosigmoid polyps and 171 small non-rectosigmoid polyps were included in the study. The NPV for discriminating neoplastic polyps in small rectosigmoid polyps with CAD-AFI was 93.4\% (184/197) [95\% confidence interval (CI), 89.0\%-96.4\%] and with TME it was 94.9\% (185/195) (95\% CI, 90.8\%-97.5\%). For differentiating tiny rectosigmoid neoplastic polyps by CAD-AFI, the accuracy, sensitivity, specificity, and positive predictive value were 91.5\%, 80.0\%, 95.3\%, and 85.2\%, respectively.

  & Small rectosigmoid polyps could be distinguished well using real-time CAD-AFI. This objective technology can aid the efficient management of tiny rectosigmoid polyps without advanced training.

\\
 (Steenhuis et al., 2020)  \cite{steenhuis2020feasibility}      & Feasibility of volatile organic compound in breath analysis in the follow-up of colorectal cancer: A pilot study

 &The author investigated whether the AeonoseTM eNose under investigation can identify local recurrence or metastasis of CRC.

  &62 patients in this cross-sectional analysis had received curative treatment for CRC within the previous five years. 26 of them had extraluminal local recurrence or metastases of CRC found during FU, while 36 had no metastases. Breath tests and machine learning were utilised to predict extraluminal recurrences or metastases, and sensitivity and specificity were estimated based on the receiver operating characteristic (ROC)-curve

  &  With a sensitivity and specificity of 0.88 (CI 0.69-0.97) and 0.75 (CI 0.57-0.87), respectively, and an overall accuracy of 0.81, the eNose detected extra luminal local recurrences or metastases of CRC.

& This eNose might help find extraluminal local recurrences or metastases in the FU of CRCs that have had curative treatment. Before it may be utilised in clinical practice, a well-designed prospective study is required to demonstrate its accuracy and predictive value

\\
  (Schrammen et al., 2022)  \cite{schrammen2022weakly}     & Weakly supervised annotation-free cancer detection and prediction of genotype in routine histopathology

 & innovative technique for detecting tumours and predicting genetic changes at the same time: The Slide-Level Assessment Model (SLAM) improves upon earlier approaches by automatically removing regular and non-informative tissue sections to predict molecular changes straight from routine pathology slides without any operator annotations

 &  Using two sizable multicentric colorectal cancer patient populations, Darmkrebs: Chancen der Verhütung durch Screening (DACHS) from Germany and Yorkshire Cancer Research Bowel Cancer Improvement Programme (YCR-BCIP) from the UK,  have thoroughly verified SLAM for therapeutically relevant tasks.
& real under the receiver operating curve (AUROC) of 0.980 (confidence interval 0.975, 0.984; n = 2,297 tumour and n = 1,281 normal slides), SLAM produces accurate slide-level classification of tumour presence. Additionally, SLAM can identify BRAF mutation status with an AUROC of 0.821 (0.786, 0.852; n = 2,075 patients) and microsatellite instability (MSI)/mismatch repair deficiency (dMMR) or microsatellite stability/mismatch repair proficiency. An extensive external testing cohort was used to validate the improvement over earlier techniques, and MSI/dMMR status was identified with an AUROC of 0.900 (0.864, 0.931; n = 805 individuals)

 & SLAM offers visualisation maps that are easy for humans to understand, making it possible for specialists to analyse multiplexed network forecasts. In conclusion, SLAM is a novel, straightforward, and effective computational pathology technique that may be used in a variety of disease scenarios

\\
  (Vleugels et al., 2019) \cite{vleugels2019diminutive}      &  Diminutive Polyps With Advanced Histologic Features Do Not Increase Risk for Metachronous Advanced Colon Neoplasia

& To know the percentage of patients who are at high risk for metachronous advanced neoplasia due to small polyps and advanced histologic characteristics in colonoscopy

 & Collected information from 12 cohorts of patients (in the United States or Europe) who had colonoscopies performed after receiving a positive faecal immunochemical test result (FIT cohort, n = 34,221) or who had colonoscopies performed for screening, surveillance, or symptom evaluation (colonoscopy cohort, n = 30,123)

 & Patients with polyps that exhibited advanced histologic signs (cancer, high-grade dysplasia, 25\% villous features), 3 or more tiny or small (6-9 mm) nonadvanced adenomas, or an adenoma or sessile serrated lesion 10 mm, were at high risk for metachronous advanced neoplasia. Calculated the proportion of diminutive polyps with advanced histologic features, the percentage of patients categorised as high risk because their diminutive polyps had advanced histologic features, and the risk of these patients for metachronous advanced neoplasia using an inverse variance random effects model.
However, the prevalence difference did not result in a significant difference in the proportions of patients assigned to high-risk status (0.8\% of patients in the FIT cohort and 0.4\% of patients in the colonoscopy cohort) (P =.25). In 51,510 diminutive polyps, advanced histologic features were observed in 7.1\% of polyps from the FIT cohort and 1.5\% of polyps from the colonoscopy cohort (P =.044). The proportion of low-risk patients with metachronous advanced neoplasia (14.6\%) did not differ significantly from the proportion of high-risk patients (17.6\%) due to diminutive polyps with advanced histologic features (relative risk for high-risk categorisation, 1.13; 95\% confidence interval 0.79-1.61)
 & tiny polyps with advanced histologic features do not raise risk for metachronous advanced neoplasia in a pooled study of data from 12 international cohorts of individuals undergoing colonoscopy for screening, surveillance, or evaluation of symptoms

\\
     (Brockmoeller et al., 2022) \cite{brockmoeller2022deep}    & Deep learning to identify inflamed fat as a risk factor for lymph node metastasis in early colorectal cancer

 & To understand the molecular mechanisms underlying thepoorly-known occurrence of colorectal cancer, early-stage (T1 and T2) adenocarcinomas that have migrated to local lymph nodes are a crucial occurrence (CRC), and the prognostic biomarkers that are currently available are not ideal. 

 & Digitised histology slides of the main CRC and its surrounding tissue to determine risk variables for lymph node metastasis (LNM) status using an end-to-end deep learning method

 & The results indicated several LNMs in pT2 CRC patients with an area under the receiver operating curve (AUROC) of 0.733 (0.67-0.758) and patients with any LNM with an AUROC of 0.711 in two sizable population-based datasets (0.597–0.797). The occurrence of many LNMs or any LNM was also expected in pT1 CRC patients, with an AUROC of 0.733 (0.644-0.778) and 0.567 (0.542-0.597), respectively

 & The method can accurately predict the presence of several LNMs in pT2 CRC patients with an area

\\
     (Backes et al., 2019) \cite{backes2019multicentre}   & Multicentre prospective evaluation of real-time optical diagnosis of T1 colorectal cancer in large non-pedunculated colorectal polyps using narrow-band imaging (the OPTICAL study)

 &massive non-pedunculated colorectal polyps were used to assess the preresection accuracy of optical diagnosis of T1 colorectal cancer (CRC) (LNPCPs)

  & Endoscopists used a standardised method for optical assessment to predict the histology during colonoscopy in consecutive individuals with LNPCPs. Along with the optical diagnosis, the degree of prediction confidence, and the suggested course of therapy, the presence of morphological features examined under white light, vascular patterns assessed under narrow-band imaging (NBI), and surface patterns analysed under NBI were all recorded. A multivariable mixed effects binary logistic least absolute shrinkage and selection (LASSO) model was used to create and evaluate a risk score table.

 & 47 malignancies (36 T1 CRCs and 11 T2 CRCs) were discovered among 343 LNPCPs, of which 11 T1 CRCs were superficial invasive T1 CRCs (23.4\% of all malignant polyps). The sensitivity and specificity for the optical diagnosis of T1 CRC were 78.7\% (95\% CI 64.3 to 89.3) and 94.2\% (95\% CI 90.9 to 96.6), and 63.3\% (95\% CI 43.9 to 80.1) and 99.0\% (95\% CI 97.1 to 100.0) and respectively for the optical diagnosis of endoscopically unresectable lesions (i.e., T1 CRC with deep invasion). A cross-validation area under the curve (AUC) of 0.85 (95\% CI 0.80 to 0.90) distinguishes T1 CRCs from non-invasive polyps in a LASSO-derived model employing white light and NBI. A temporal validation set of 100 LNPCPs was used to validate this model (AUC of 0.81; 95\% CI 0.66 to 0.96).

 & Further research will demonstrate how risk score charts could be enhanced and ultimately used for clinical decision-making regarding the endoresection employed and whether to proceed with surgery rather than endoscopy. Sensitivity is currently restricted.

\\
    \label{tab:my_label}
\end{longtable}

\subsection{Appendix B}
\subsubsection{Abbreviations}
\begin{abbrv}

\item[BPPS]			Boston Bowel Preparation Scale

\item[CADe]			Computer-Aided Detection
\item[CCE]			Colon Capsule Endoscopies
\item[CTC]			Computed Tomography Colonography 
\item[OC]			Optical Colonoscopy
\item[CONSORT]			Consolidated Standards of Reporting Trials
\item[CRC]			Colorectal Cancer
\item[DS-STM]			Diagnosis Strategy of Serum Tumor Makers

\item[DA]			Diagnosis Accuracy
\item[HDWL]			High-Definition White Light

\item[PDR]			Polyp Detection Rate
\item[PMR]			Polyp Miss Rate

\item[QUADAS-2]			Quality Assessment of Diagnostic Accuracy Studies 
\item[NGS]			Next Generation Sequencing

\item[SPIRIT]			Standard Protocol Items: Recommendations for Interventional Trials

\item[SSL]			Sessile Serrated Lesions

\item[SSLPC]			Sessile Serrated Lesions Per Colonoscopy

\item[U.S.]			United States

\end{abbrv}
\end{document}